\documentclass[12pt]{article}
\usepackage{epsf}
\title{ {\bf
Lepton flavor violating radion decays in the Randall-Sundrum
scenario}}
\author{\vspace{1cm}\\
        {\bf E. O. Iltan,}
        \thanks{E-mail address:
        eiltan@newton.physics.metu.edu.tr}
 {\,\bf B. Korutlu}
        \thanks{E-mail address:
        beste.korutlu@gmail.com}
 \\
        Physics Department, Middle East Technical University \\
        Ankara, Turkey\\}

\date{}

\begin{document}
\setlength{\baselineskip}{24pt}
\maketitle
\setlength{\baselineskip}{7mm}
\begin{abstract}
We predict the branching ratios of the lepton flavor violating
radion decays $r\rightarrow e^{\pm} \mu^{\pm}$, $r\rightarrow
e^{\pm} \tau^{\pm}$ and $r\rightarrow \mu^{\pm} \tau^{\pm}$ in the
two Higgs doublet model, in the framework of the Randall-Sundrum
scenario. We observe that their branching ratios are at most of
the order of $10^{-8}$, for the small values of radion mass and
they decrease with the increasing values of $m_r$. Among these
processes, the $r\rightarrow \tau^{\pm}\, \mu^{\pm}$ decay would
be the most suitable one to measure its branching ratio.
\end{abstract}
\thispagestyle{empty}
\newpage
\setcounter{page}{1}
\section{Introduction}
The hierarchy problem between weak and Planck scales could be
explained by introducing the extra dimensions. One of the
possibility is to pull down the Planck scale to TeV range by
considering the compactified extra dimensions of large size
\cite{ArkaniHamed}. The assumption that the extra dimensions are
at the order of submilimeter distance, for two extra dimensions,
the hierarchy problem in the fundamental scales could be solved
and the true scale of quantum gravity would be no more the Planck
scale but it is of the order of electroweak (EW) scale. This is
the case that the gravity spreads over all the volume including
the extra dimensions, however, the matter fields are restricted in
four dimensions, so called four dimensional (4D) brane. Another
possibility, which is based on the non-factorizable geometry, is
introduced by Randall and Sundrum \cite{Rs1} (the RS1 model) and,
in this scenario, the extra dimension is compactified to $S^1/Z_2$
orbifold with two 4D brane boundaries. Here, the gravity is
localized in one of the boundary, so called the Planck (hidden)
brane, which is away from another boundary, the TeV (visible)
brane where we live. The size of extra dimension is related to the
vacuum expectation of a scalar field and its fluctuation over the
expectation value is called the radion field (see section 2 for
details). The radion in the RS1 model has been studied in several
works in the literature \cite{Goldberger}-\cite{Kingman} (see
\cite{Csaba} for extensive discussion).

In the present work, we study the possible lepton flavor violating
(LFV) decays of the radion field $r$. The LFV interactions exist
at least in one loop level in the extended standard model (SM), so
called $\nu$SM, which is constructed by taking neutrinos massive
and by permitting the lepton mixing mechanism \cite{Pontecorvo}.
Their negligibly small branching ratios (BRs) stimulate one to go
beyond and they are worthwhile to examine since they open a window
to test new models and to ensure considerable information about
the restrictions of the free parameters, with the help of the
possible accurate measurements. The LFV interactions are carried
by the flavor changing neutral currents (FCNCs) and in the SM with
extended Higgs sector (the multi Higgs doublet model) they can
exist at tree level. Among multi Higgs doublet models, the two
Higgs doublet model (2HDM) is a candidate for the lepton flavor
violation. In this model, the lepton flavor violation is driven by
the new scalar Higgs bosons $S$, scalar $h^0$ and pseudo scalar
$A^0$, and it is controlled by the Yukawa couplings appearing in
lepton-lepton-S vertices.

Here, we predict the BRs of the LFV $r$ decays in the 2HDM, in the
framework of the RS1 scenario \footnote{The flavor conserving
decays of the radion can exist in the tree level and the
theoretical values of the BRs are free from the loop suppressions.
For the heavy flavor output $\tau\tau$ the BR reaches to the
numerical values of the order of $10^{-2}$ for the small values of
the radion mass, $m_r< 200\, GeV$. In the case of light flavor
output $\mu\mu \, (ee)$ the numerical value of the BR is of the
order of $10^{-4}$ ($10^{-9}$) for $m_r< 200\, GeV$ (see
\cite{Cheung} for the explicit expressions of these decay
widths).}. The BRs of these processes are sensitive to the new
Yukawa couplings arising with the addition of the new Higgs
doublet and they enhance with the increasing values of the these
couplings. Therefore, besides their important role in the
construction of the FCNC in the tree level and the existence of
the LFV interactions, these Yukawa couplings play a crucial role
in the enhancement of the theoretical values of the BRs. In our
calculations, we observe that the BRs of the processes we study
are at most of the order of $10^{-8}$, for the small values of
radion mass $m_r$ and for the large values of the Yukawa
couplings, and, their sensitivities to $m_r$ decrease with the
increasing values of $m_r$. Among the LFV decays we study, the
$r\rightarrow \tau^{\pm}\, \mu^{\pm}$ decay would be the most
suitable one to measure its BR.

The paper is organized as follows: In Section 2, we present the
effective vertex and the BRs of LFV $r$ decays in the 2HDM, by
respecting the RS1 scenario. Section 3 is devoted to a discussion
and to our conclusions. In the appendix, we present the
interaction vertices appearing in the calculations.
\section{The LFV RS1 radion decay in the 2HDM}
The RS1 model is an interesting candidate in order to explain the
well known hierarchy problem. It is formulated as two 4D surfaces
(branes) in 5D world in which the extra dimension is compactified
into $S^1/Z_2$ orbifold. In this model, the SM fields are assumed
to live on one of the brane, so called the TeV brane. On the other
hand, the gravity peaks near the other brane, so called the Planck
brane and extends into the bulk with varying strength. Here, 5D
cosmological constant is non vanishing and both branes have equal
and opposite tensions so that the low energy effective theory has
flat 4D spacetime. The metric of such 5D world reads
\begin{eqnarray}
ds^2=e^{-2\,A(y)}\,\eta_{\mu\nu}\,dx^\mu\,dx^\nu-dy^2\, ,
\label{metric1}
\end{eqnarray}
where $A(y)=k\,|y|$, k is the bulk curvature constant, y is the
extra dimension parametrized as $y=R\,\theta$. The exponential
factor $e^{-k\,L}$ with $L=R\,\pi$, is the warp factor which
causes that all the mass terms are rescaled in the TeV brane. With
a rough estimate  $L\sim 30/k$, all mass terms are brought down to
the TeV scale. The size $L$ of extra dimension is related to the
vacuum expectation of the field $L(x)$ and its fluctuation over
the expectation value is called the radion field $r$. In order to
avoid the violation of equivalence principle, $L(x)$ should
acquire a mass and, to stabilize $r$, a mechanism was proposed by
Goldberger and Wise \cite{Goldberger}, by introducing a potential
for $L(x)$. Finally, the metric in 5D is defined as \cite{Rubakov}
\begin{eqnarray}
ds^2=e^{-2\,A(y)-2\,F(x)}\,\eta_{\mu\nu}\,dx^\mu\,dx^\nu- (1+2\,
F(x))\,dy^2\, , \label{metric2}
\end{eqnarray}
where the radial fluctuations are carried by the scalar field
$F(x)$,
\begin{eqnarray}
F(x)=\frac{1}{\sqrt{6}\,M_{Pl}\,e^{-k\,L}}\, r(x)\, . \label{Fx}
\end{eqnarray}
Here the field $r(x)$ is the normalized radion field (see
\cite{CsabaMichael}). At the orbifold point $\theta=\pi$ (TeV
brane) the induced metric reads,
\begin{eqnarray}
g^{ind}_{\mu\nu}=e^{-2\,A(L)-2\frac{\gamma}{v}\,r(x)}\,\eta_{\mu\nu}
\, . \label{metricind}
\end{eqnarray}
Here the parameter $\gamma$ reads
$\gamma=\frac{v}{\sqrt{6}\,\Lambda}$ with
$\Lambda=M_{Pl}\,e^{-k\,L}$ and $v$ is the vacuum expectation
value of the SM Higgs boson. The radion is the additional degree
of freedom of the 4D effective theory and we study the possible
LFV decays of this field.

The FCNCs at tree level can exist in the 2HDM and they induce the
flavor violating (FV) interactions with large BRs.  The FV $r$
decays, $r\rightarrow l_1^- l_2^+$, can exist at least in one loop
level in the framework of the 2HDM. The part of action which
carries the interaction, responsible for the LFV processes reads
\begin{eqnarray}
{\cal{S}}_{Y}&=& \int d^4x \sqrt{-g^{ind}}\Bigg(
\eta^{E}_{ij} \bar{l}_{i L} \phi_{1} E_{j R}+ \xi^{E}_{ij}
\bar{l}_{i L} \phi_{2} E_{j R} + h.c. \Bigg) \,\,\, ,
\label{lagrangian1}
\end{eqnarray}
where $L$ and $R$ denote chiral projections $L(R)=1/2(1\mp
\gamma_5)$, $\phi_{i}$ for $i=1,2$, are two scalar doublets, $l_{i
L}$ ($E_{j R}$) are lepton doublets (singlets), $\xi^{E}_{ij}$
\footnote{In the following, we replace $\xi^{E}$ with
$\xi^{E}_{N}$ where "N" denotes the word "neutral".} and
$\eta^{E}_{ij}$, with family indices $i,j$ , are the Yukawa
couplings and $\xi^{E}_{ij}$ induce the FV interactions in the
leptonic sector. Here $g^{ind}$ is the determinant of the induced
metric on the TeV brane where the 2HDM particles live. Here, we
assume that the Higgs doublet $\phi_1$ has a non-zero vacuum
expectation value to ensure the ordinary masses of the gauge
fields and the fermions, however, the second doublet has no vacuum
expectation value, namely, we choose the doublets $\phi_{1}$ and
$\phi_{2}$ and their vacuum expectation values as
\begin{eqnarray}
\phi_{1}=\frac{1}{\sqrt{2}}\left[\left(\begin{array}{c c}
0\\v+H^{0}\end{array}\right)\; + \left(\begin{array}{c c} \sqrt{2}
\chi^{+}\\ i \chi^{0}\end{array}\right) \right]\, ;
\phi_{2}=\frac{1}{\sqrt{2}}\left(\begin{array}{c c} \sqrt{2}
H^{+}\\ H_1+i H_2 \end{array}\right) \,\, , \label{choice}
\end{eqnarray}
and
\begin{eqnarray}
<\phi_{1}>=\frac{1}{\sqrt{2}}\left(\begin{array}{c c}
0\\v\end{array}\right) \,  \, ; <\phi_{2}>=0 \,\, .\label{choice2}
\end{eqnarray}
This choice ensures that the mixing between neutral scalar Higgs
bosons is switched off and it would be possible to separate the
particle spectrum so that the SM particles are collected in the
first doublet and the new particles in the second one
\footnote{Here $H_1$ ($H_2$) is the well known mass eigenstate
$h^0$ ($A^0$).}. The action in eq. (\ref{lagrangian1}) is
responsible for the tree level $S-l_1-l_2$ ($l_1$ and $l_2$ are
different flavors of charged leptons, $S$ denotes the neutral new
Higgs boson, $S=h^0,A^0$) interaction (see Fig. 1-d, e) and the
four point $r-S-l_1-l_2$ interaction (see Fig. 1-c) where $r$ is
the radion field. The latter interaction is coming from the
determinant factor
$\sqrt{-g^{ind}}=e^{-4\,A(L)-4\frac{\gamma}{v}\,r(x)}$. Notice
that the term $e^{-4\,A(L)}$ in $\sqrt{-g^{ind}}$ is embedded into
the redefinitions of the fields on the TeV brane, namely, they are
warped as $S\rightarrow e^{A(L)}\,S_{warp}$, $l\rightarrow
e^{3\,A(L)/2}\,l_{warp}$ and in the following we use warped fields
without the $warp$ index.

On the other hand, the part of new scalar action
\begin{eqnarray}
{\cal{S}}_{2}&=& \int d^4x \sqrt{-g^{ind}} \Bigg( g^{ind\,\,
\mu\nu}\, (D_\mu\,\phi_2)^\dagger \, D_\nu\, \phi_2-m_S^2\,
\phi_2^\dagger \,\phi_2\Bigg)  \label{lagrangian2}
\end{eqnarray}
leads to
\begin{eqnarray}
{\cal{S'}}_{2}\!\!\!\!&=&\!\! \!\!\frac{1}{2}\,\int d^4x \,\Bigg\{
e^{-2\frac{\gamma}{v}\, r}\eta^{\mu\nu}\,
\Big(\partial_\mu\,h^0\partial_\nu\,h^0+\partial_\mu\,A^0\partial_\nu\,A^0
\Big)-e^{-4\frac{\gamma}{v}\,
r}\,(m_{h^0}^2\,h^0\,h^0+m_{A^0}^2\,A^0\,A^0)\Bigg\} \, ,
\label{lagrangian2a}
\end{eqnarray}
which carries the $S-S-r$ interaction\footnote{In general, there
is no symmetry which forbids the curvature-scalar interaction,
\begin{eqnarray}
{\cal{S}}_{\xi}=\int d^4x
\,\sqrt{-g^{ind}}\,\xi\,{\cal{R}}\,H^\dagger\,H \, ,
\label{curvscalarmix}
\end{eqnarray}
where $\xi$ is a restricted positive parameter and $H$ is the
Higgs scalar field \cite{CsabaMichael,Kingman,Csaba}. This
interaction results in the radion-(SM or new) Higgs mixing which
can bring a sizeable contribution to the physical quantities
studied. Here, we assume that there is no mixing between first and
second doublet and only the first Higgs doublet has vacuum
expectation value. Therefore, we choose that there exists a mixing
between the radion and the SM Higgs field, but not between the
radion and the new Higgs fields. This is the case that the lepton
flavor violation is not affected by the mixing since the SM Higgs
field is not responsible for the FCNC current at tree level.} (see
Fig. 1-b).

Finally, the interaction of leptons with the radion field is
carried by the action (see \cite{Kumar})
\begin{eqnarray}
{\cal{S}}_{3}&=& \int d^4x \sqrt{-g^{ind}}\, \Bigg( g^{ind\,\,
\mu\nu}\, \bar{l}\,\gamma_\mu\,i\,D_\nu\,l-m_l\,\bar{l}\,l\Bigg)
\, , \label{lagrangian3}
\end{eqnarray}
where
\begin{eqnarray}
D_\mu \,l=\partial_\mu\,l+\frac{1}{2}\,w_\mu^{ab}\,\Sigma_{ab}\,l
\, , \label{Dmuf}
\end{eqnarray}
with $\Sigma_{ab}=\frac{1}{4}[\gamma_a,\gamma_b]$. Here
$w_\mu^{ab}$ is the spin connection and, by using the vierbein
fields $e^a_\mu$, it can be calculated (linear in $r$) as
\begin{eqnarray}
w_\mu^{ab}=-\frac{\gamma}{v}\partial_\nu\,r\,(e^{\nu
b}\,e^a_\mu-e^{\nu a}\,e_\mu^b )\, . \label{vierbein}
\end{eqnarray}
Notice that the vierbein fields are the square root of the metric
and they satisfy the relation
\begin{eqnarray}
e^\mu_a\,e^{a \nu}=g^{ind\,\, \mu\nu}\, . \label{metrvierbein}
\end{eqnarray}
Using eqs. (\ref{lagrangian3})-(\ref{metrvierbein}), one gets the
part of the action which describes the tree level $l-l-r$
interaction  (see Fig.1-a) as
\begin{eqnarray}
{\cal{S'}}_{3}&=&\int d^4x \,\Bigg\{ -3\frac{\gamma}{v}\, r\,
\bar{l}\,i\,\partial\!\!\!/ l-3\frac{\gamma}{2\,v}\,
\bar{l}\,i\,\partial\!\!\! / r\, l+ 4\,\frac{\gamma}{v}\, m_l\,r\,
\bar{l}\,l \Bigg\} \, . \label{lagrangian3a}
\end{eqnarray}

At this stage, we would like to discuss briefly the case that the
gauge fields and the leptons are also accessible to the extra
dimension. With the addition of Dirac mass term
$m_D(y)=m\frac{A'(y)}{k}$ ($A'(y)=\frac{dA(y)}{dy}$) to the
lagrangian of bulk fermions, the fermion mass hierarchy can be
obtained since they have different locations, regulated by a
localization parameter $c=m/k$, in the extra dimension (see for
example \cite{Pamoral2, Grossman, Huber, Huber2}). Furthermore,
the SM fermions are obtained by considering $SU(2)_L$ doublet
$\psi_L$ and singlet $\psi_R$ with separate $Z_2$ projection
conditions: $Z_2\psi_R=\gamma_5 \psi_R$ and $Z_2\psi_L=-\gamma_5
\psi_L$ (see for example \cite{Hisano}) and the localization
parameters $c_L$ and $c_R$ assigned to each left and right handed
fermion. If $c_{L;R}>\frac{1}{2}$ ($c_{L;R}<\frac{1}{2}$), the
zero mode left;right handed fermions are localized near the hidden
(visible) brane and, for $c_{L;R}=\frac{1}{2}$, they have constant
profiles. Notice that the parameters $c_{L}$ and $c_{R}$ should be
restricted such that the observed masses of fermions are obtained.
In addition to the SM fermions (the zero mode ones) the KK modes
of left and right handed fermions arise and these modes depend on
the localization parameters $c_{L}$ and $c_{R}$. In this scenario,
the BRs of the LFV radion decays become sensitive to the different
locations of left and right handed (zero mode and KK mode)
leptons. This sensitivity is due to the couplings coming from
$r-l^{(n)}_{iL(R)}-l^{(n)}_{iR (L)}$,
$r-l^{(n)}_{iL(R)}-l^{(n)}_{iL(R)}$ and
$S-l^{(n)}_{iL(R)}-l^{(n)}_{iR(L)}$ interactions, where $S=h^0,
A^0$ and $n=0,1,2,...$. Here we can take two different
possibilities: The first possibility is that  the existence of the
LFV is based on the different locations of different flavors (see
for example \cite{KAgashe, EBlechman, Ledroit}). In this case the
new Yukawa couplings, appearing with the interaction of zero mode
leptons and new Higgs bosons, sensitive to the locations of the
corresponding leptons and the sources of the flavor violation are
explicitly the localization parameters. If the leptons are far
from (near to) the visible brane the couplings are suppressed
(enhanced). Furthermore, the strengths of the couplings of KK mode
leptons and new Higgs bosons are regulated by the locations of the
lepton fields. In the second possibility, the flavor violation is
carried by the new Yukawa couplings in four dimensions by assuming
that the localization effects are embedded into the definition of
new Yukawa couplings and, with the additional effects coming from
the KK mode leptons, the role of extra dimension becomes the
enhancement in the physical quantities . In both scenarios, the
couplings of the radion with the leptons are regulated by the
localization of the zero mode leptons and there exists an
additional effect coming from the radion-KK mode lepton vertices.
At first sight, it is expected that if the zero leptons are far
from (near to) the visible brane these couplings are suppressed
(enhanced). However, one needs a detailed analysis to take into
account the bulk lepton contributions and their localization
effects.

Now, we are ready to calculate the matrix element for the LFV
radion decay. The decay of the radion to leptons with different
flavors exits at least in one loop order, with the help of
internal new Higgs bosons $S=h^0, A^0$. The possible vertex and
self energy diagrams are presented in Fig. \ref{figselfvert}.
After addition of all these diagrams, the divergences which occur
in the loop integrals are eliminated and the matrix element square
for this decay is obtained as
\begin{eqnarray}
|M|^2= 2\Big( m_r^2 -(m_{l_1^-}+m_{l_2^+})^2\Big)\,|A|^2 \, ,
\label{Matrx2}
\end{eqnarray}
where
\begin{eqnarray}
A= f^{self}_{h^0}+f^{self}_{A^0}+f^{vert}_{h^0}+f^{vert}_{A^0}+
f^{vert}_{h^0 h^0}+f^{vert}_{A^0 A^0} \, , \label{fh0A0}
\end{eqnarray}
and the explicit expressions of the functions appearing in eq.
(\ref{fh0A0}) are given in Appendix B. Notice that in eq.
(\ref{etaVA}) the flavor changing couplings $\xi^{E}_{N, l_ji}$
represent the effective interaction between the internal lepton
$i$, ($i=e,\mu,\tau$) and the outgoing $j=1\,(j=2)$ lepton (anti
lepton). Here, we choose the couplings $\xi^{E}_{N, l_ji}$ real.

Finally, the BR for $r\rightarrow l_1^-\,l_2^+$ can be obtained by
using the matrix element square as
\begin{eqnarray}
BR (r\rightarrow l_1^- \,l_2^+)=\frac{1}{16\,\pi\,m_r}\,
\frac{|M|^2}{\Gamma_r}\, , \label{BR1}
\end{eqnarray}
where $\Gamma_r$ is the total decay width of radion $r$. In our
numerical analysis,  we consider the BR due to the production of
sum of charged states, namely
\begin{eqnarray}
BR (r\rightarrow l_1^{\pm}\,l_2^{\pm})= \frac{\Gamma(r\rightarrow
(\bar{l}_1\,l_2+\bar{l}_2\,l_1))}{\Gamma_r} \, .\label{BR2}
\end{eqnarray}
%
\section{Discussion}
In four dimensions, the higher dimensional gravity is observed as
it has new states with spin 2,1 and 0, so called, the graviton,
the gravivector, the graviscalar. These states interact with the
particles in the underlying theory. In the RS1 model with one
extra dimension, the spin 0 gravity particle radion $r$ interacts
with the particles of the theory (2HDM in our case) on the TeV
brane and this interaction occurs over the trace of the
energy-momentum tensor $T^\mu_\mu$ with the strength
$1/\Lambda_r$,
\begin{eqnarray}
{\cal{L}}_{int}=\frac{r}{\Lambda_r}\,T^\mu_\mu \, ,\label{Radint}
\end{eqnarray}
where $\Lambda_r$  is at the order of TeV. The radion interacts
with gluon ($g$) pair or photon ($\gamma$) pair in one loop order
from the trace anomaly. For the radion mass $m_r\le 150\,GeV$, the
decay width is dominated by $r\rightarrow gg$.  For the masses
which are beyond the WW and ZZ thresholds, the main decay mode is
$r\rightarrow WW$. In the present work, we study the possible LFV
decays of the RS1 radion in the 2HDM and estimate the BRs of these
decays for different values of radion masses. We take the total
decay width $\Gamma_r$ of the radion by considering the dominant
decays $r\rightarrow gg\, (\gamma\gamma,  ff, W^+W^-,ZZ, SS)$
where $S$ are the neutral Higgs particles (see \cite{Cheung} for
the explicit expressions of these decay widths). Here, we include
the possible processes in the $\Gamma_r$ according to the mass of
the radion.

The flavor violating  $r$ decays $r\rightarrow l_1^- l_2^+$ can
exist at least in one loop level, in the framework of the 2HDM and
the flavor violation is carried by the Yukawa couplings
$\bar{\xi}^{E}_{N,ij}$\footnote{The dimensionfull Yukawa couplings
$\bar{\xi}^{E}_{N,ij}$ are defined as
$\xi^{E}_{N,ij}=\sqrt{\frac{4\,G_F}{\sqrt{2}}}\,
\bar{\xi}^{E}_{N,ij}$.}. In the version of 2HDM where the FCNC are
permitted, these couplings are free parameters which should be
restricted by using the present and forthcoming experiments. At
first, we assume that these couplings are symmetric with respect
to the flavor indices $i$ and $j$. Furthermore, we take that the
couplings which contain at least one $\tau$ index are dominant and
we choose a broad range for these couplings, by respecting the
upper limit prediction of $\bar{\xi}^{E}_{N,\tau \mu}$ (see
\cite{Iltananomuon} and references therein), which is obtained by
using the experimental uncertainty, $10^{-9}$, in the measurement
of the muon anomalous magnetic moment and by assuming that the new
physics effects can not exceed this uncertainty. For the coupling
$\bar{\xi}^{E}_{N,\tau e}$, the restriction is estimated by using
this upper limit and the experimental upper bound of BR of
$\mu\rightarrow e \gamma$ decay, BR $\leq 1.2\times 10^{-11}$
\cite{Brooks}. Finally, this coupling is taken in the range
$10^{-3}-10^{-1}\, GeV$ (see \cite{Iltan1}). For the Yukawa
coupling $\bar{\xi}^{E}_{N,\tau \tau}$, we have no explicit
restriction region and we use the numerical values which are
greater than $\bar{\xi}^{E}_{N,\tau \mu}$.
%
%
Throughout our calculations we use the input values given in Table
(\ref{input}).
\begin{table}[h]
        \begin{center}
        \begin{tabular}{|l|l|}
        \hline
        \multicolumn{1}{|c|}{Parameter} &
                \multicolumn{1}{|c|}{Value}     \\
        \hline \hline
        $m_{\mu}$                   & $0.106$ (GeV) \\
        $m_{\tau}$                  & $1.78$ (GeV) \\
        $m_{h^0}$           & $100$   (GeV)  \\
        $m_{A^0}$           & $200$   (GeV)  \\
        $G_F$             & $1.16637 10^{-5} (GeV^{-2})$  \\
        \hline
        \end{tabular}
        \end{center}
\caption{The values of the input parameters used in the numerical
          calculations.}
\label{input}
\end{table}

In Fig.\ref{RtotaumumR} we present  $m_r$ dependence of the BR
$(r\rightarrow \tau^{\pm}\, \mu^{\pm})$. The solid-dashed lines
represent the BR $(r\rightarrow \tau^{\pm}\, \mu^{\pm})$ for
$\bar{\xi}^{E}_{N,\tau \tau}=100\,GeV$, $\bar{\xi}^{E}_{N,\tau
\mu}=10\,GeV$- $\bar{\xi}^{E}_{N,\tau \tau}=10\,GeV$,
$\bar{\xi}^{E}_{N,\tau \mu}=1\,GeV$. It is observed that the BR
$(r\rightarrow \tau^{\pm}\, \mu^{\pm})$ is of the order of the
magnitude of $10^{-8}$ for the large values of the couplings and
the radion mass values $\sim 200\,GeV$. For the heavy masses of
the radion the BR is stabilized to the values of the order of
$10^{-9}$.

Fig.\ref{RtomuetauemR} is devoted to $m_r$ dependence of the BR
$(r\rightarrow \tau^{\pm}\, e^{\pm})$ and BR $(r\rightarrow
\mu^{\pm}\, e^{\pm})$. The solid-dashed lines represent the BR
$(r\rightarrow \tau^{\pm}\, e^{\pm})$ for $\bar{\xi}^{E}_{N,\tau
\tau}=100\,GeV$, $\bar{\xi}^{E}_{N,\tau e}=0.1\,GeV$-
$\bar{\xi}^{E}_{N,\tau \tau}=10\,GeV$, $\bar{\xi}^{E}_{N,\tau
e}=0.1\,GeV$. The small dashed line represents the BR
$(r\rightarrow \mu^{\pm}\, e^{\pm})$ for $\bar{\xi}^{E}_{N,\tau
\mu}=1\,GeV$, $\bar{\xi}^{E}_{N,\tau e}=0.1\,GeV$. This figure
shows that the BR $(r\rightarrow \tau^{\pm}\, \mu^{\pm})$ is of
the order of the magnitude of $10^{-12}$ for the large values of
the couplings and the radion mass values $\sim 200\,GeV$. For the
heavy masses of the radion, this BR reaches to the values less
than $10^{-14}$. The BR $(r\rightarrow \mu^{\pm}\, e^{\pm})$ is of
the order of $10^{-15}$ for $m_r\sim 200\,GeV$ and for the
intermediate values of Yukawa couplings. These BRs, especially BR
($r\rightarrow \mu^{\pm}\, e^{\pm}$), are negligibly small.

Now, we present the Yukawa coupling dependencies of the BRs of the
decays under consideration, for different radion masses .

Fig.\ref{Rtotaumuxi} represents  the $\bar{\xi}^{E}_{N,\tau \tau}$
dependence of the BR $(r\rightarrow \tau^{\pm}\, \mu^{\pm})$ for
$\bar{\xi}^{E}_{N,\tau \mu}=10 \,GeV$. The solid-dashed-small
dashed lines represent the BR for the radion masses
$m_r=200-500-1000\, GeV$. This figure shows that the BR is
sensitive to the radion mass and, obviously, it is enhanced two
orders of magnitude in the range $10\,GeV\leq
\bar{\xi}^{E}_{N,\tau \tau}\leq 100\,GeV$.

In Fig.\ref{Rtotauexi}, we present the $\bar{\xi}^{E}_{N,\tau
\tau}$ dependence of the BR $(r\rightarrow \tau^{\pm}\, e^{\pm})$
for $\bar{\xi}^{E}_{N,\tau e}=0.1 \,GeV$. The solid-dashed-small
dashed lines represent the BR for the radion masses
$m_r=200-500-1000\,GeV$. Similar to the $r\rightarrow \tau^{\pm}\,
\mu^{\pm}$ decay, the BR is strongly sensitive to the radion mass.
%
\section{Conclusion}
The LFV decays of the radion in the RS1 model strongly depend on
the radion mass and the Yukawa couplings. The BR for $r\rightarrow
\tau^{\pm}\, \mu^{\pm}$ decay is of the order of $10^{-8}$ for the
small values of radion mass $m_r$ and it decreases with the
increasing values of $m_r$. On the other hand, the BRs for
$r\rightarrow \tau^{\pm}\, e^{\pm}$ $(r\rightarrow \mu^{\pm}\,
e^{\pm})$ decays are of the order of $10^{-12}$ ($10^{-15}$) for
the small values of $m_r$. These results show that, among these
processes, the LFV $r\rightarrow \tau^{\pm}\, \mu^{\pm}$ decay
would be the most appropriate one to measure its BR. The most
probable production of radion is due to the gluon fusion,
$gg\rightarrow r$ \cite{Cheung}) and, based on the existing cross
section values for the radion production via gluon fusion and the
typical integrated luminosities expected at LHC, a rough
estimation of the number of events for the $\tau^{\pm}\,
\mu^{\pm}$ final state is missing at present. The goal being to
check that this number of events is significant statistically, and
hopefully, the forthcoming experimental results of this decay
would be useful in order to test the possible signals coming from
the extra dimensions and new physics which results in flavor
violation.
\appendix
\section{The vertices appearing in the present work}
In this section we present the vertices appearing in our
calculations. Here $S$ denotes the new neutral Higgs bosons $h^0$
and $A^0$.
\begin{figure}
\begin{tabular}{p{6cm} p{5cm}} \vskip -7.9truein
\parbox[b]{16cm}{\epsffile{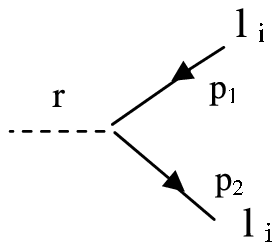}} &
\raisebox{5.ex}{$\frac{-i\,\gamma}{v}\,\left[\frac{3}{2}\,
(p_1\!\!\!\!\!/+p_2\!\!\!\!\!/\,)-4\,m_{l_i}\right]$}
\\ \textbf{(a)} \\ \\ \vskip -7.0truein
\parbox[b]{6cm}{\epsffile{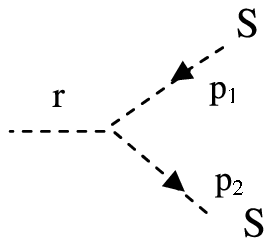}} & \\  \vskip 0.2truein \hspace{6.3cm}
\raisebox{5.ex} {$\frac{-2\,i\,\gamma}{v}\, (p_1.p_2-m_S^2)$}
\\ \textbf{(b)}\\ \\ \vskip -7.1truein
\parbox[b]{6cm}{\epsffile{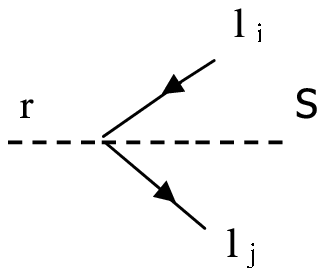}} & \vspace{0.7cm}\\ \vskip -0.4truein
\hspace{4.7cm}
\raisebox{5.ex}{$(S=h^0)\,\,\,\,\,\,\,\frac{4\,i\gamma}{2\sqrt{2}\,v}
\left[(\xi_{ij}^{E}+
\xi_{ji}^{E*})+(\xi_{ij}^{E}-\xi_{ji}^{E*})\gamma_5\right]$}
\vspace{0.7cm}\\ \vskip -0.5truein \hspace{4.7cm}
\raisebox{5.ex}{$(S=A^0)\,\,\,\,\,\,\,\frac{-4\,\gamma}{2\sqrt{2}\,v}\left[
(\xi_{ij}^{E}-\xi_{ji}^{E*})+(\xi_{ij}^{E}+\xi_{ji}^{E*})\gamma_5\right]$}
\textbf{(c)}\\ \\ \\ \vskip -7.3truein
\parbox[b]{6cm}{\epsffile{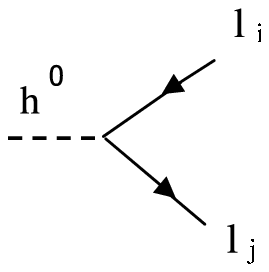}} & \vspace{0.5cm}\\ \hspace{6.3cm}
\raisebox{5.ex}{$\frac{-i}{2\sqrt{2}}\left[(\xi_{ij}^{E}+
\xi_{ji}^{E*})+(\xi_{ij}^{E}-\xi_{ji}^{E*})\gamma_5\right]$}
\vspace{-1.5cm} \textbf{(d)} \\ \\
\\ \\ \\ \\
\vskip -7.7truein
\parbox[b]{6cm}{\epsffile{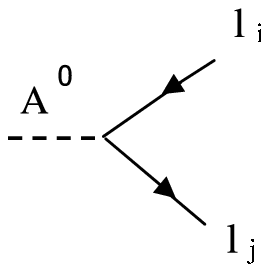}} & \vspace{-1.5cm}\\ \vskip -0.4truein
\hspace{6.3cm}
\raisebox{5.ex}{$\frac{1}{2\sqrt{2}}\left[(\xi_{ij}^{E}-
\xi_{ji}^{E*})+(\xi_{ij}^{E}+\xi_{ji}^{E*})\gamma_5\right]$}  \textbf{(e)}\\ \\
\\ \\ \\ \\
\end{tabular}
%
\label{figvert1} \caption{The vertices used in the present work.}
\end{figure}
\newpage
\section{The explicit expressions appearing in the text}
The explicit expressions of functions $f^{self}_{h^0\,(A^0)}$,
$f^{ver}_{h^0\,(A^0)}$, $f^{ver}_{h^0\,h^0}$ and
$f^{ver}_{A^0\,A^0}$ in eq.(\ref{fh0A0}) are given by
\begin{eqnarray}
f^{self}_{h^0}&=& \frac{\gamma}{128\,v\,\pi^2\,(w'_h-w_h) }
\int_0^1\,dx\, m_{h^0}\,
 \Bigg \{\Big( \eta_i^V\,(x-1)\,w_h-\eta_i^+\,z_{ih} \Big)\, \Big(
3\,w'_h-5\,w_h \Big)\,\,ln\,\frac{L^{self}_{1,
h^0}\,m^2_{h^0}}{\mu^2}\nonumber \\
&+& \Big( \eta_i^V\,(x-1)\,w'_h-\eta_i^+\,z_{ih} \Big)\, \Big(
5\,w'_h-3\,w_h \Big)\,\,ln\,\frac{L^{self}_{2,
h^0}\,m^2_{h^0}}{\mu^2}\Bigg \} \, , \nonumber \\
f^{self}_{A^0}&=& \frac{\gamma}{128\,v\,\pi^2\,(w'_A-w_A) }
\int_0^1\,dx\ ,m_{A^0}\,
 \Bigg \{\Big( \eta_i^V\,(x-1)\,w_A+\eta_i^+\,z_{iA} \Big)\, \Big(
3\,w'_A-5\,w_A \Big)\,\,ln\,\frac{L^{self}_{1,
A^0}\,m^2_{A^0}}{\mu^2}\nonumber \\
&+& \Big( \eta_i^V\,(x-1)\,w'_A+\eta_i^+\,z_{iA} \Big)\, \Big(
5\,w'_A-3\,w_A \Big)\,\,ln\,\frac{L^{self}_{2,
A^0}\,m^2_{A^0}}{\mu^2}\Bigg \} \, , \nonumber \\
f^{vert}_{h^0}&=& \frac{\gamma}{128\,v\,\pi^2} \int_0^1\,dx\,
\int_0^{1-x} \, dy \, \Bigg\{
\frac{m_{h^0}}{L^{ver}_{h^0}}\,\Bigg[ \eta_i^V
\Bigg( 3\, z_{rh}^2\,\Big(y\,(1-y)\,w'_h+x^2\,(4\,y-1)\,w_h \nonumber \\
&+& x\,((1-3\,y)\,w_h+y\,(4\,y-3)\,w'_h)\Big)
+ 5\, z_i^2\, \Big( (2\,y-1)\,w'_h+(2\,x-1)\,w_h \Big)\nonumber
\\&-&
3\, (x+y-1)\,\Big(x\,(4\,x-3)\,w_h^3+y\,(4\,y-3)\,w_h^{'3}\Big) \nonumber \\
&-& 3\, w_h\, w'_h \,(x+y-1)\,\Big(
 (1-y+x\,(4\,y-2))\,w_h+  (1-2\,y+x\,(4\,y-1))\,w'_h \Big)\nonumber \\
&+& 3\, (x+y-1)\,\Big((2\,x-1)\,w_h+(2\,y-1)\,w'_h \Big)
 \Bigg)\nonumber \\
&+& \eta_i^+ \,z_{ih}\,\Bigg(
(x+y-1)\,\Big(-4+2\,w'_h\,w_h+w_h^{'2}\,(8\,y-3)+w_h^{2}\,(8\,x-3)\,\Big)
\nonumber \\
&-& \Big(8\,z_{ih}^2+z_{rh}^2\,((8\,y-3)\,x-3\,y)\Big)\,\Bigg)
\Bigg]
\nonumber \\
&-& m_{h^0}\,ln\,\frac{L^{ver}_{h^0}\,m^2_{h^0}}{\mu^2}\,\Bigg(
9\, \eta_i^V \Big(w'_h\,(2\,y-1)+w_h\,(2\,x-1)\Big)- 8\,\eta_i^+
\,z_{ih} \Bigg)
\Bigg \}\, , \nonumber \\
f^{vert}_{A^0}&=& \frac{\gamma}{128\,v\,\pi^2} \int_0^1\,dx\,
\int_0^{1-x} \, dy \, \Bigg\{
\frac{m_{A^0}}{L^{ver}_{A^0}}\,\Bigg[ \eta_i^V
\Bigg( 3\, z_{rA}^2\,\Big(y\,(1-y)\,w'_A+x^2\,(4\,y-1)\,w_A \nonumber \\
&+& x\,((1-3\,y)\,w_A+y\,(4\,y-3)\,w'_A)\Big)
+ 5\, z_i^2\, \Big( (2\,y-1)\,w'_A+(2\,x-1)\,w_A \Big)\nonumber
\\&-&
3\, (x+y-1)\,\Big(x\,(4\,x-3)\,w_A^3+y\,(4\,y-3)\,w_A^{'3}\Big) \nonumber \\
&-& 3\, w_A\, w'_A \,(x+y-1)\,\Big(
 (1-y+x\,(4\,y-2))\,w_A+  (1-2\,y+x\,(4\,y-1))\,w'_A \Big)\nonumber \\
&+& 3\, (x+y-1)\,\Big((2\,x-1)\,w_A+(2\,y-1)\,w'_A \Big)
 \Bigg)\nonumber \\
&+& \eta_i^+ \,z_{iA}\,\Bigg(
(x+y-1)\,\Big(-4+2\,w'_A\,w_A+w_A^{'2}\,(8\,y-3)+w_A^{2}\,(8\,x-3)\,\Big)
\nonumber \\
&-& \Big(8\,z_{iA}^2+z_{rA}^2\,((8\,y-3)\,x-3\,y)\Big)\,\Bigg)
\Bigg]
\nonumber \\
&-& m_{A^0}\,ln\,\frac{L^{ver}_{A^0}\,m^2_{A^0}}{\mu^2}\,\Bigg(
9\, \eta_i^V \Big(w'_A\,(2\,y-1)+w_A\,(2\,x-1)\Big)+ 8\,\eta_i^+
\,z_{iA} \Bigg)
\Bigg \}\, , \nonumber \\
f^{vert}_{h^0\,h^0}&=& \frac{\gamma}{64\,v\,\pi^2} \int_0^1\,dx\,
\int_0^{1-x} \, dy \, \Bigg \{
\frac{m_{h^0}}{L^{ver}_{h^0\,h^0}}\,\Bigg[ \eta_i^V
\Bigg( z_{rh}^2\,\Big( y-1+x\,(1-4\,y)\Big)\,(x\,w_h+y \,w'_h) \nonumber \\
&+&
y\,(x+y-1)\,w'_h\,\Big( (4\,x-1)\,w_h^2+(4\,y-1)\,w_h^{'2} \Big) \nonumber \\
&+& w_h^3\,x\,(x+y-1)\,(4\,x-1)+
(x+y-1)\,\Big(2\,y\,w'_h+x\,w_h\,(2+w_h^{'2}\,(4\,y-1))\,\Big)
 \Bigg) \nonumber \\
&+& \eta_i^+ \Bigg(
(x+y-1)\,z_{ih}\,\Big((4\,y-1)\,w_h^{'2}+(4\,x-1)\,w_h^2+2\Big)-
z_{ih}\,z_{rh}^2\,\Big((4\,y-1)\,x-y+1\Big)\Bigg)\,\nonumber
\Bigg] \nonumber \\
&-&
m_{h^0}\,ln\,\frac{L^{ver}_{h^0\,h^0}\,m^2_{h^0}}{\mu^2}\,\Bigg(
\, \eta_i^V \Big(w'_h\,(1-6\,y)+w_h\,(1-6\,x)\Big)- 4\,\eta_i^+
\,z_{ih} \Bigg)
\Bigg \}\, , \nonumber \\
f^{vert}_{A^0\,A^0}&=& \frac{\gamma}{64\,v\,\pi^2} \int_0^1\,dx\,
\int_0^{1-x} \, dy \, \Bigg \{
\frac{m_{A^0}}{L^{ver}_{A^0\,A^0}}\,\Bigg[ \eta_i^V
\Bigg( z_{rA}^2\,\Big( y-1+x\,(1-4\,y)\Big)\,(x\,w_A+y \,w'_A) \nonumber \\
&+&
y\,(x+y-1)\,w'_A\,\Big( (4\,x-1)\,w_A^2+(4\,y-1)\,w_A^{'2} \Big) \nonumber \\
&+& w_A^3\,x\,(x+y-1)\,(4\,x-1)+
(x+y-1)\,\Big(2\,y\,w'_A+x\,w_A\,(2+w_A^{'2}\,(4\,y-1))\,\Big)
 \Bigg) \nonumber \\
&+& \eta_i^+ \Bigg(
(x+y-1)\,z_{iA}\,\Big((4\,y-1)\,w_A^{'2}+(4\,x-1)\,w_A^2+2\Big)-
z_{iA}\,z_{rA}^2\,\Big((4\,y-1)\,x-y+1\Big)\Bigg)\,\nonumber
\Bigg] \nonumber \\
&-&
m_{A^0}\,ln\,\frac{L^{ver}_{A^0\,A^0}\,m^2_{A^0}}{\mu^2}\,\Bigg(
\, \eta_i^V \Big(w'_A\,(1-6\,y)+w_A\,(1-6\,x)\Big)+ 4\,\eta_i^+
\,z_{iA} \Bigg)
\Bigg \}\, , \label{fVAME}
\end{eqnarray}
where
\begin{eqnarray}
L^{self}_{1, h^0 (A^0)}&=&1+x^2\,w_{h(A)}^{2}+x\,(z^2_{ih(iA)}-
w_{h(A)}^{2}-1)\nonumber \, , \\
L^{self}_{2, h^0 (A^0)}&=&1+x^2\,w_{h(A)}^{'2}+x\,(z^2_{ih
(iA)}-w_{h(A)}^{'2}-1)
\nonumber \, , \\
L^{ver}_{h^0
(A^0)}&=&x^2\,w_{h(A)}^{2}+(y-1)\,(w_{h(A)}^{'2}\,y-1)+x\,(y\,w_{h(A)}^{'2}+
(y-1)\,w_{h(A)}^{2}-y\,z^2_{rh (rA)}-1)
\nonumber \, , \\
L^{ver}_{h^0\,h^0
(A^0\,A^0)}\!\!\!\!&=&\!\!\!\!x^2\,w_{h(A)}^{2}+(1+w_{h(A)}^{'2}\,(y-1))\,y+
x\,(1+w_{h(A)}^{2}\,(y-1)+w_{h(A)}^{'2}\,y-z^2_{rh (rA)}\,y) \, ,
\nonumber \\ \label{Lh0A0}
\end{eqnarray}
with the parameters $w_{h(A)}=\frac{m_{l_1^-}}{m_{h^0 (A^0)}}$,
$w'_{h(A)}=\frac{m_{l_2^+}}{m_{h^0 (A^0)}}$, $z_{rh
(rA)}=\frac{m_r}{m_{h^0(A^0)}}$, $z_{ih
(iA)}=\frac{m_i}{m_{h^0(A^0)}}$ and
\begin{eqnarray}
\eta_i^V&=&\xi^{E}_{N,l_li}\xi^{E\,*}_{N,il_2}+
\xi^{E\,*}_{N,il_1} \xi^{E}_{N,l_2 i} \nonumber \, , \\
\eta_i^+&=&\xi^{E\,*}_{N,il_1}\xi^{E\,*}_{N,il_2}+
\xi^{E}_{N,l_1i} \xi^{E}_{N,l_2 i} \, . \label{etaVA}
\end{eqnarray}
%
%
\newpage
\newpage
\begin{figure}[htb]
\vskip 2.0truein \epsfxsize=5.5in
\leavevmode\epsffile{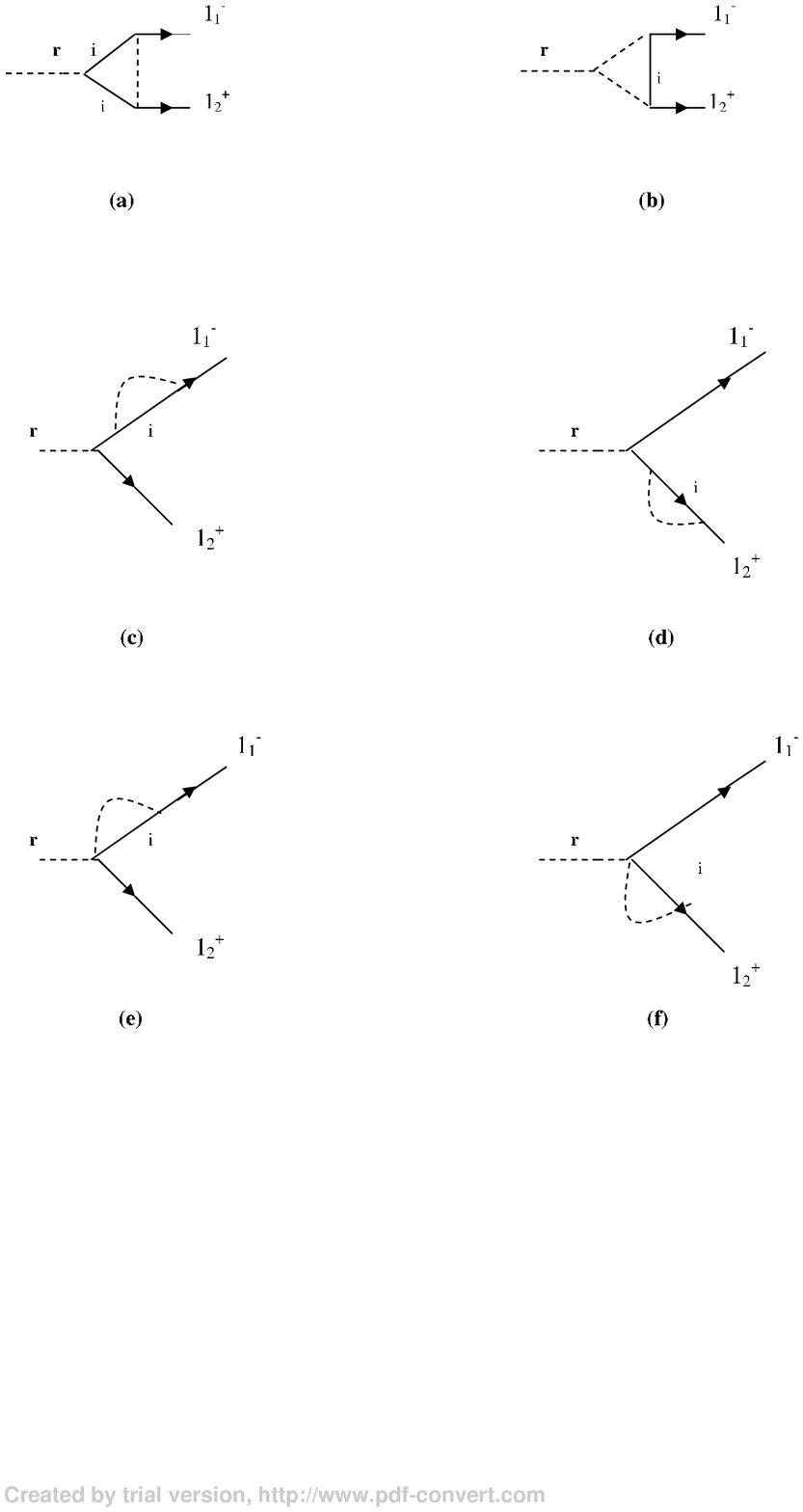} \vskip -3.9truein
\caption[]{One loop diagrams contribute to $r\rightarrow
l_1^-\,l_2^+$ decay due to the neutral Higgs bosons $h_0$ and
$A_0$ in the 2HDM. $i$ represents the internal lepton, $l_1^-$
($l_2^+$) outgoing lepton (anti lepton), internal dashed line the
$h_0$ and $A_0$ fields.} \label{figselfvert}
\end{figure}
\newpage
\begin{figure}[htb]
\vskip -3.0truein \centering \epsfxsize=6.8in
\leavevmode\epsffile{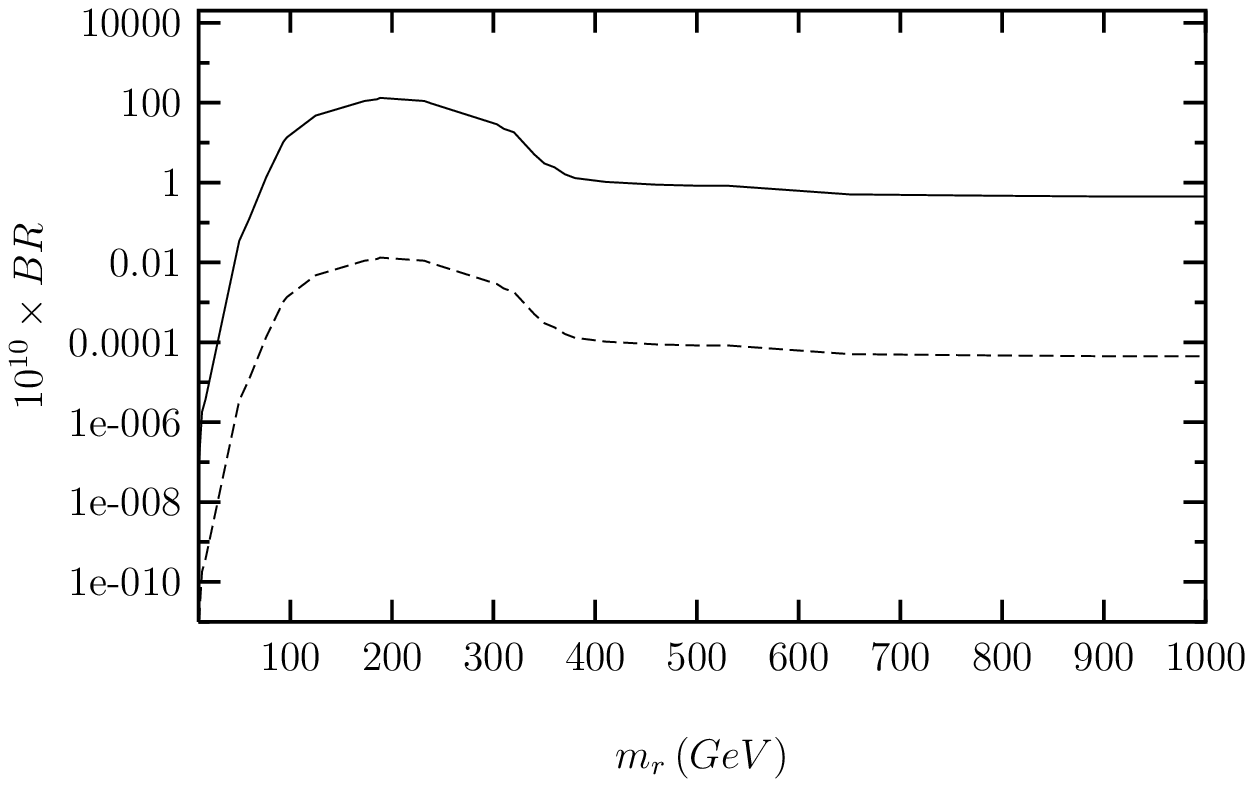} \vskip -3.0truein \caption[]{
$m_r$ dependence of the BR $(r\rightarrow \tau^{\pm}\,
\mu^{\pm})$. The solid-dashed lines represent the BR$(r\rightarrow
\tau^{\pm}\, \mu^{\pm})$ for $\bar{\xi}^{E}_{N,\tau
\tau}=100\,GeV$, $\bar{\xi}^{E}_{N,\tau \mu}=10\,GeV$-
$\bar{\xi}^{E}_{N,\tau \tau}=10\,GeV$, $\bar{\xi}^{E}_{N,\tau
\mu}=1\,GeV$.} \label{RtotaumumR}
\end{figure}
\begin{figure}[htb]
\vskip -3.0truein \centering \epsfxsize=6.8in
\leavevmode\epsffile{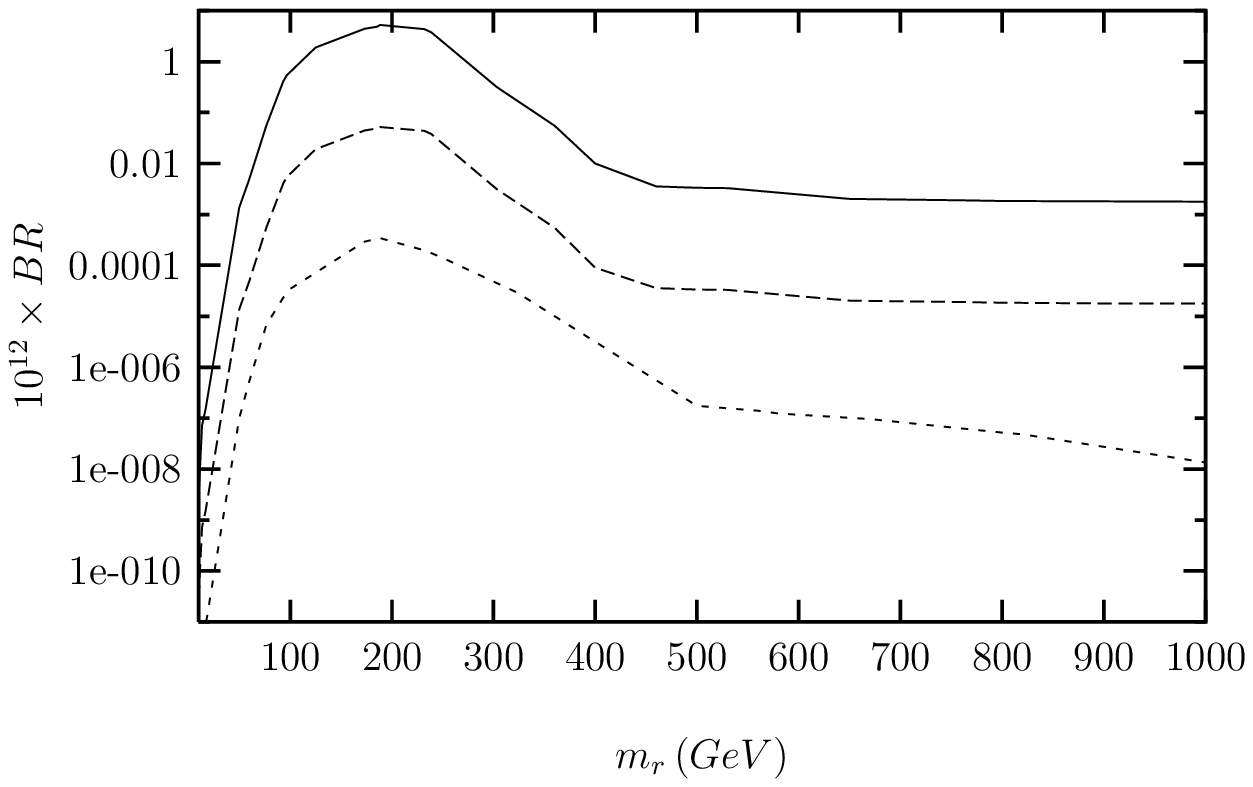} \vskip -3.0truein
\caption[]{ $m_r$ dependence of the BR $(r\rightarrow \l_1^{\pm}\,
l_2^{\pm})$. The solid-dashed lines represent the BR$(r\rightarrow
\tau^{\pm}\, e^{\pm})$ for $\bar{\xi}^{E}_{N,\tau \tau}=100\,GeV$,
$\bar{\xi}^{E}_{N,\tau e}=0.1\,GeV$- $\bar{\xi}^{E}_{N,\tau
\tau}=10\,GeV$, $\bar{\xi}^{E}_{N,\tau e}=0.1\,GeV$. The small
dashed line represents the BR $(r\rightarrow \mu^{\pm}\, e^{\pm})$
for $\bar{\xi}^{E}_{N,\tau \mu}=1\,GeV$, $\bar{\xi}^{E}_{N,\tau
e}=0.1\,GeV$.} \label{RtomuetauemR}
\end{figure}
\begin{figure}[htb]
\vskip -3.0truein \centering \epsfxsize=6.8in
\leavevmode\epsffile{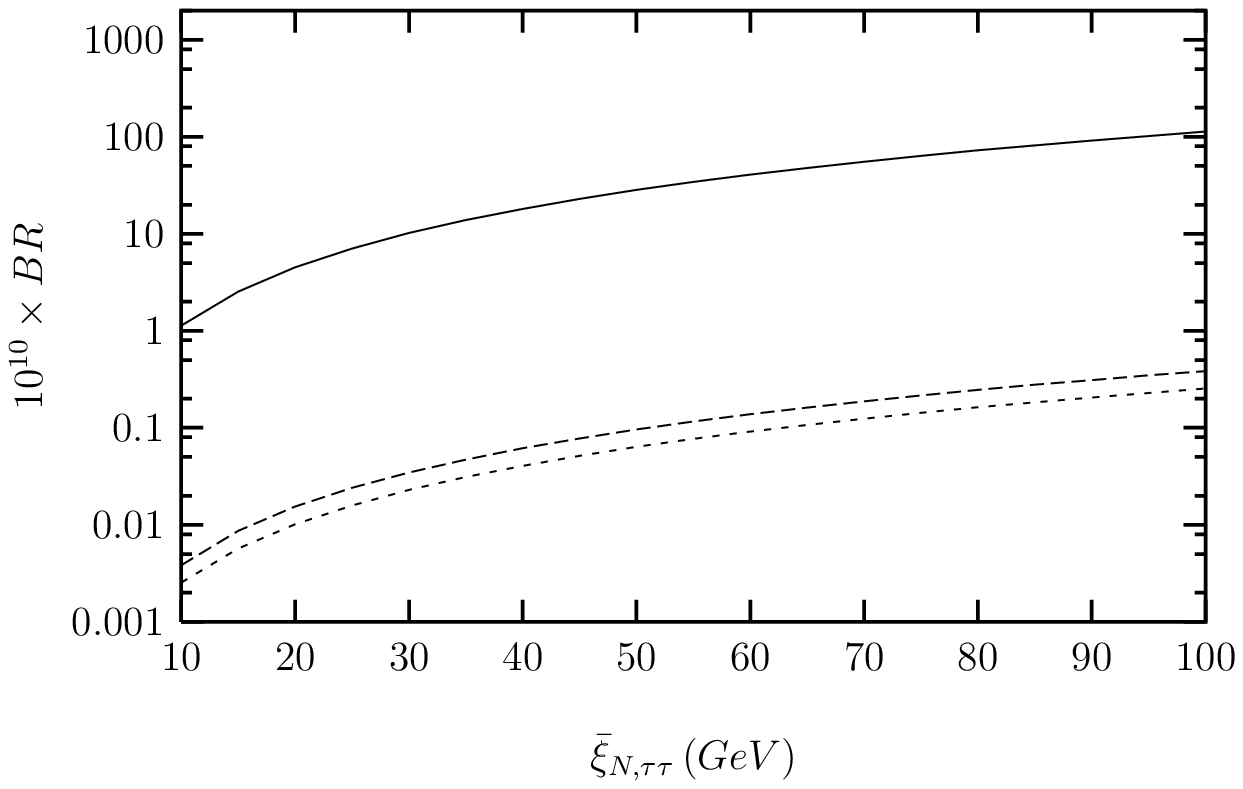} \vskip -3.0truein \caption[]{
$\bar{\xi}^{E}_{N,\tau \tau}$ dependence of the BR$(r\rightarrow
\tau^{\pm}\, \mu^{\pm})$ for $\bar{\xi}^{E}_{N,\tau \mu}=10
\,GeV$. The solid-dashed-small dashed lines represent the BR for
the radion masses $m_r=200-500-1000\,GeV$. } \label{Rtotaumuxi}
\end{figure}
\begin{figure}[htb]
\vskip -3.0truein \centering \epsfxsize=6.8in
\leavevmode\epsffile{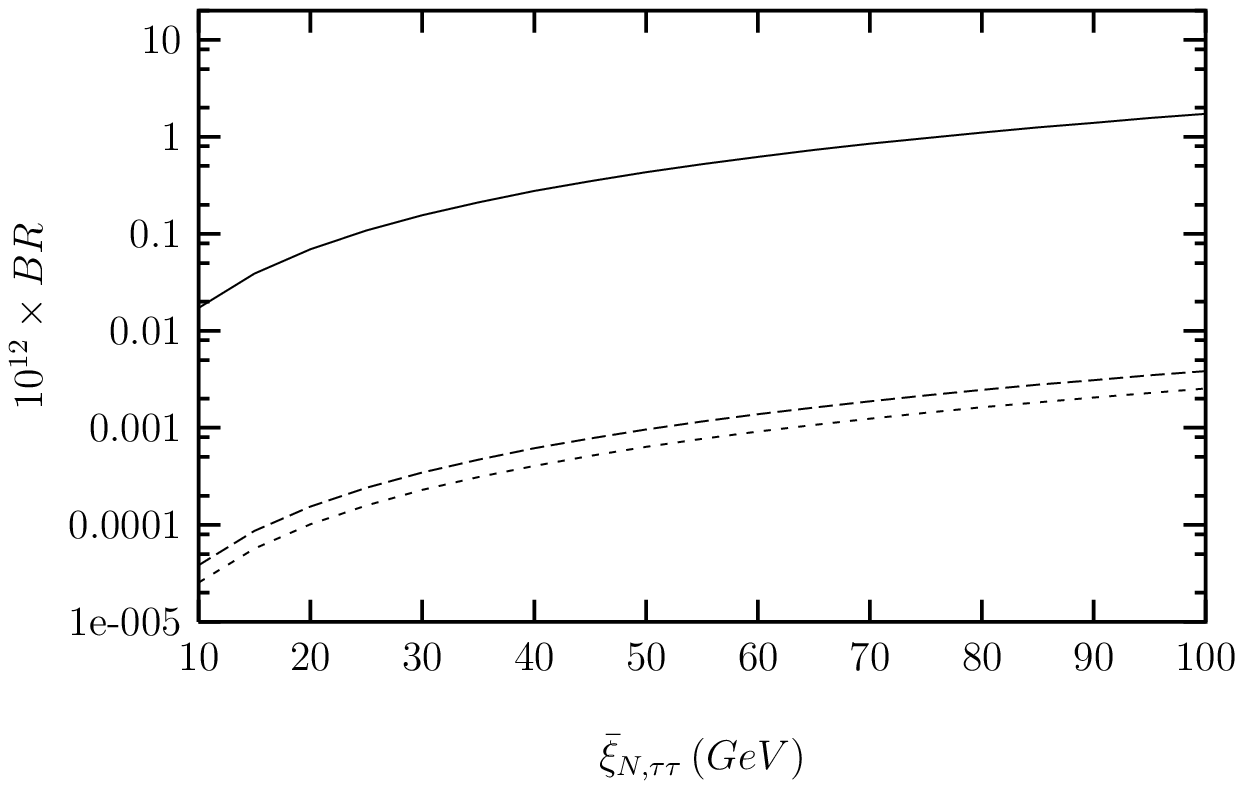} \vskip -3.0truein
\caption[]{$\bar{\xi}^{E}_{N,\tau \tau}$ dependence of the
BR$(r\rightarrow \tau^{\pm}\, e^{\pm})$ for $\bar{\xi}^{E}_{N,\tau
e}=0.1 \,GeV$. The solid-dashed-small dashed lines represent the
BR for the radion masses $m_r=200-500-1000\,GeV$.}
\label{Rtotauexi}
\end{figure}
\end{document}